\documentclass[prb, twocolumn,amsmath,amssymb]{revtex4}
\usepackage{graphicx, color}
\usepackage{dcolumn}
\usepackage{bm}
\usepackage{epstopdf}
\usepackage{xcolor}
\usepackage{subfigure}
\definecolor{TTH-color}{named}{green}
\definecolor{PV-color}{rgb}{0.97,0.57,0.11}
\definecolor{YL-color}{RGB}{128,128,0}
\definecolor{TTH-color2}{named}{red}
\definecolor{PV-color2}{rgb}{0.87,0.47,0.01}
\definecolor{YL-color2}{RGB}{128,128,0}

\begin{document}
\title{Effect of disorder on Majorana localization in topological superconductors: a quasiclassical approach} 

\author{Yao Lu}
\author{Pauli Virtanen}
\author{Tero T. Heikkil\"a}

\affiliation{Department of Physics and Nanoscience Center, University of Jyv\"askyl\"a,
P.O. Box 35 (YFL), FI-40014 University of Jyv\"askyl\"a, Finland}
\date{\today}
\begin{abstract}
Two dimensional topological superconductors (TS) host chiral Majorana modes (MMs) localized at the boundaries. In this work, within quasiclassical approximation we study the effect of disorder on the localization length of MMs in two dimensional spin-orbit (SO) coupled superconductors. We find nonmonotonic behavior of the Majorana localization length as a function of disorder strength. At weak disorder, the Majorana localization length decreases with an increasing disorder strength. Decreasing the disorder scattering time below a critical value $\tau_c$, the Majorana localization length starts to increase.  The critical scattering time depends on the relative magnitudes of the two ingredients behind TS: SO coupling and exchange field. For dominating SO coupling, $\tau_c$ is small and vice versa for the dominating exchange field.

\end{abstract}

\maketitle

\section{Introduction}

Realization of topological superconductors (TSs) supporting Majorana modes (MMs) in condensed matter systems has attracted much attention due to its potential application in quantum computing [\onlinecite{qi2011topological,kitaev2001unpaired,fu2008superconducting,sau2010generic,yamakage2012evolution,law2009majorana,mourik2012signatures,nadj2014observation,he2017chiral}]. As random impurities are variantly present in any realistic systems, understanding the effect of disorder on the Majorana localization length is of great importance and interest. It was commonly believed that unlike $s$ wave superconductors, topological superconductors should be treated as effective unconventional superconductors (like $p$ wave superconductors) which violate Anderson's theorem and are very sensitive to disorder. MMs cannot survive when the disorder strength is much larger than the pairing gap, in which case the bulk spectrum becomes gapless.

Plenty of works have been devoted to study the effect of disorder on MMS in one dimensional $p$ wave superconductors [\onlinecite{motrunich2001griffiths,brouwer2011probability,lobos2012interplay,pientka2013signatures,huse2013localization,adagideli2014effects,hui2014generalized,stanev2014}]. It has been shown that disorder reduces the bulk energy gap and increases the localization length of MMs. A phase transition to a topologically trivial phase occurs at the gap closing point where the localization length of MMs diverges.  For multi-channel systems [\onlinecite{potter2010multichannel,rieder2013reentrant,rieder2014density,lu2016influence,burset2017current}], the behavior is similar to the single channel case at weak disorder, but can go through multiple phase transitions at stronger disorder.

Recently, it has been reported that in planar Josephson junctions which are effectively one-dimensional TSs [\onlinecite{hell2017two,pientka2017topological,hell2017coupling,hart2017controlled,glodzik2020measure}] weak disorder can also decrease the Majorana localization length [\onlinecite{haim2019benefits}]. The low energy physics can be described by a one-dimensional multiple-channel model.  In this model, different channels experience different pairing potentials and the Majorana localization length is determined by the pairing potential with the smallest magnitude. The effect of disorder is to average the pairing potential between the channels. Thus the smallest pairing potential 
increases and the Majorana localization length decreases. 

Two dimensional TS supporting chiral Majorana edge modes were theoretically proposed [\onlinecite{qi2010chiral,chung2011conductance,wang2015chiral,tanaka1,tanaka2,tanaka3,tanaka4,tanaka5}] and experimentally realized [\onlinecite{he2017chiral}] in a quantum anomalous Hall insulator-superconductor structure. However, we are not aware of a previous study on the effect of disorder in 2D TSs realized in SO coupled systems. Although the effect of disorder on the chiral Majorana modes has been investigated in $p$ wave superfluids/superconductors [\onlinecite{tsutsumi2012edge,suzuki2016spontaneous,bakurskiy2014anomalous,sauls2011surface}], in SO coupled systems with proximity induced $s$ wave pairing, the results should be different and depend on the ratio between SO coupling and spin-splitting strength. 

In this work, we study the properties of MMs in single band spin-orbit coupled superconductors in the presence of weak disorder. Spin-orbit coupled superconductors subjected to an external magnetic field can be driven to a topological phase and host MMs when an odd number of electron bands are partially occupied. In order to get the spatial distribution of MMs, we adopt the quasiclassical approximation, by integrating out the relative momentum in the Green function. This treatment simplifies the calculations with the price that we lose the information of the fast oscillating part of the Green function. However since we are only interested in the localization length of MMs, the fast oscillating part of the Green function is not important. At weak disorder, we analytically calculate the Majorana localization length and show that it decreases with increasing disorder strength for any SO coupling strength and exchange field. This effect of disorder is due to a renormalization of the Fermi velocity. We also numerically solve the Eilenberger equation and get the localization length for arbitrary disorder. We find that the Majorana localization length starts to increase with an increasing disorder strength when the disorder scattering time becomes shorter than a critical scattering time $\tau_c$. This critical scattering time vanishes in the strong SO limit and increases monotonically when increasing the exchange field.

\section{Model Hamiltonian}
\begin{figure}[h!]
\centering
\subfigure{\label{a}\includegraphics[width = 0.85\columnwidth]{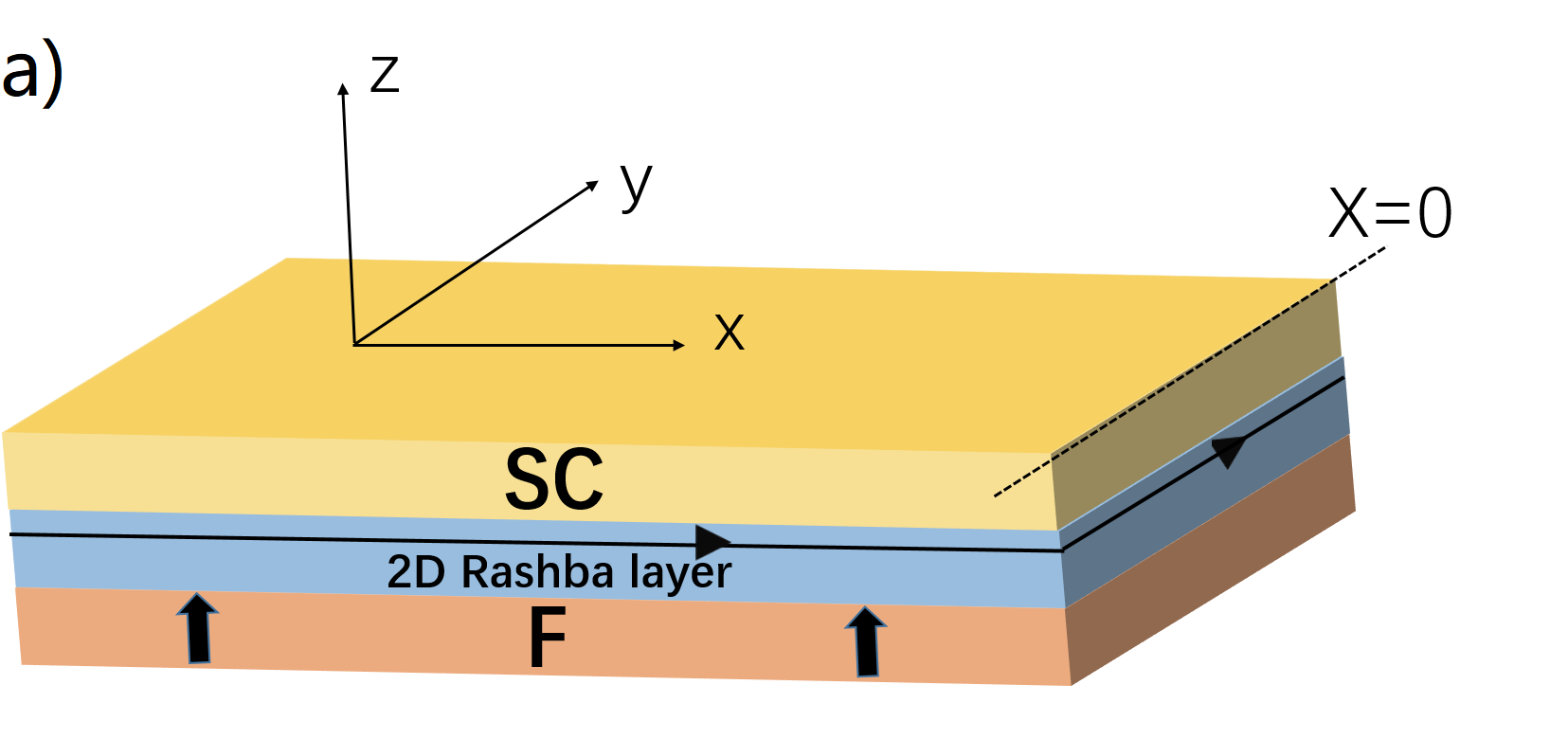}}
\subfigure{\label{b}\includegraphics[width=0.65\columnwidth]{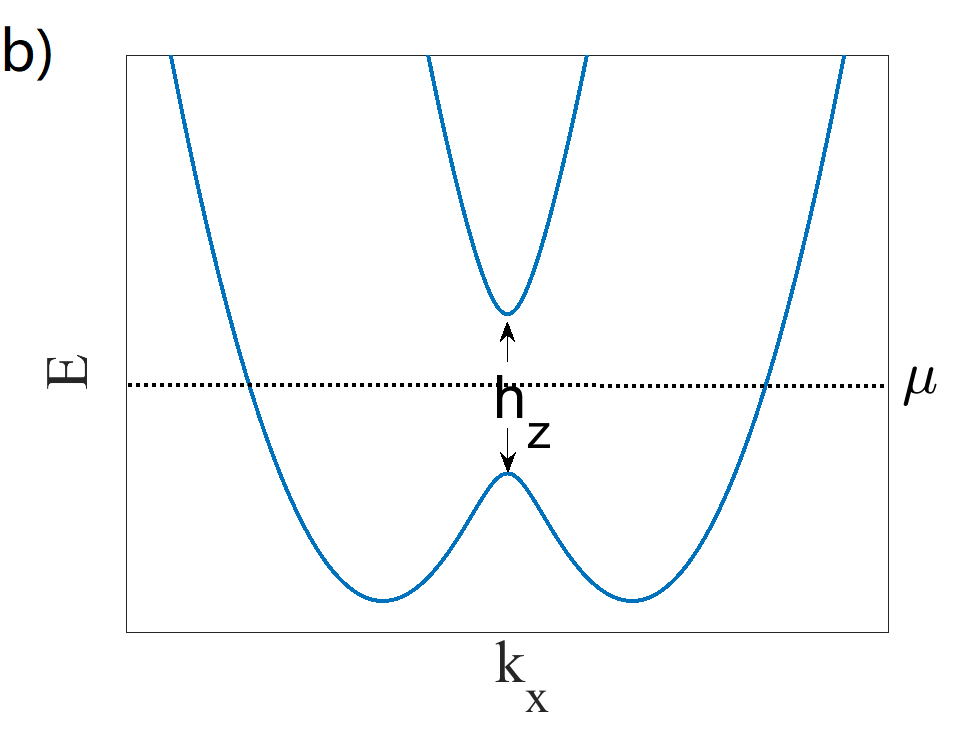}}
\caption{(a) Sketch of the system under consideration. 2D Rashba layer sandwiched by a superconducting thin film and a ferromagnetic insulator. (b) Schematic picture of the band structure. The chemical potential only cuts the lower band, so that the system is in the topological phase.}\label{Fig:setup}
\end{figure}

 We consider a heavy metal thin film with strong SO coupling sandwiched by a superconducting thin film and a ferromagnetic insulator as shown in Fig. \ref{Fig:setup}a. The effective Hamiltonian describing the 2D Rashba layer with proximity induced pairing and exchange field is given by 
 
 \begin{equation}
    H=\int d\boldsymbol{r}\quad c^{\dagger}(\boldsymbol{r})H_0c(\boldsymbol{r})+\Delta c_{\uparrow}^{\dagger}(r)c_{\downarrow}^{\dagger}(r)+h.c.,
 \end{equation}
 with
 \begin{equation}
 H_0=\frac{-\boldsymbol{\nabla}^{2}}{2m}-\mu-i\alpha(\nabla_{x}\sigma_{2}-\nabla_{y}\sigma_{1})+h_{z}\sigma_{3}+U(\boldsymbol{r}).
 \end{equation}
 
 Here $c(\boldsymbol{r})=[c_{\uparrow}(\boldsymbol{r}),c_{\downarrow}(\boldsymbol{r})]^{\text{T}}$, where $c_{s}(\boldsymbol{r})^{\dagger}$ is the creation operator which creates one electron at position $\boldsymbol{r}$ with spin $s$. $m$, $\mu$ and $\Delta$ denote the effective mass, chemical potential and pairing potential, respectively. $\alpha$ is the spin-orbit coupling coefficient and $h_{z}$ is the exchange field in the out-of-plane direction. $U(\boldsymbol{r})$ is the Gaussian disorder potential with the correlator $\langle U(\boldsymbol{r})U(\boldsymbol{r}')\rangle=\delta(\boldsymbol{r}-\boldsymbol{r}')/\pi n\tau$, where $\tau$ is the scattering time of particles in the disordered system and $n$ is the density of states per unit cell at the Fermi level. The schematic band structure (without disorder and superconductivity) is shown in Fig.~\ref{Fig:setup}b. Here we consider the case where the chemical potential only cuts the lower band, so that the system is in the topological phase and hosts chiral Majorana edge states [\onlinecite{yamakage2012evolution}]. Taking into account the effect of disorder and expressing it in spin$\otimes$particle-hole space, the Gorkov equation is given by
 
 \begin{equation}
     (G_{0}^{-1}+\mu-\hat{\Sigma})G=1,
 \end{equation}
 with
 
 \begin{equation}
   \label{eq:gorkov}
     \hat{G}_{0}^{-1}=-\frac{\boldsymbol{k}^{2}}{2m_{N}}-\left(\alpha k_{x}\sigma_{2}-\alpha k_{y}\sigma_{1}\right)+(\epsilon-h_{z}\sigma_{z})\tau_{z}.
 \end{equation}
 
 Here $\sigma_{i}$ and $\tau_i$ are Pauli matrices acting on spin and particle-hole space, respectively. $\hat{\Sigma}=\frac{\langle G\rangle}{2\tau}$ is the disorder self-energy, where $\langle\cdot\rangle$ means an average over all momenta. To investigate the properties of MMs, we assume the system is in the region $x<0$. We use periodic boundary conditions in the $y$ direction and study the Majorana edge states localized on the $x=0$ edge.
 
 \section{quasiclassical approximation}
 
 A generalized quasiclassical theory can be obtained by projecting the Gorkov Green function onto the lower band [\onlinecite{eilenberger1968transformation,lu2019proximity,zyuzin2016josephson}]. The resulting Eilenberger equation is given by (see Appendix~\ref{app:basis}) 
 
 \begin{equation}
     \boldsymbol{v}_{F}\cdot\hat{\boldsymbol{\nabla}}\hat{g}_{\boldsymbol{n}_{F}}=\left[\hat{g}_{\boldsymbol{n}_{F}},i\epsilon\tau_{3}+\Delta\tau_{1}+\hat{\Sigma}\right],\label{Eq:Eilenberger}
 \end{equation}
 where $\hat{g}_{n_{F}}$ is the quasiclassical Green function defined by
 
 \begin{equation}
     \hat{g}_{\boldsymbol{n}_{F}}(\epsilon;\boldsymbol{R})=\int\frac{d\epsilon_{p}}{i\pi}\hat{G}(\epsilon;\boldsymbol{R},\boldsymbol{p}).
 \end{equation}
 The disorder self-energy in the Born approximation becomes $\hat{\Sigma}=\frac{\langle \hat{g}\rangle_{\boldsymbol{n}_{F}}}{2\tau}$, where $\tau$ is the disorder scattering time.
 Here $\boldsymbol{n}_F$ is the unit vector along the direction of Fermi momentum $\boldsymbol{p}_F$ and $\langle\cdot\rangle_{n_{F}}$ means an angular average over all the momentum directions. This angular average should be done in the usual spin$\otimes$particle-hole space and after we get the self-energy we project it back onto the lower band sub space. The Eilenberger equation is supplemented by the normalization condition $\hat{g}^{2}=I_{P}$, where $\hat{I}_{P}$ is the identity operator in the lower band subspace. Writing $\hat{g}$ in terms of Pauli matrices $\hat{g}=g_{1}\tau_{1}+g_{2}\tau_{2}+\hat{g}_{3}\tau_{3}$, the normalization condition becomes $g_{1}^{2}+g_{2}^{2}+g_{3}^{2}=1$. In the clean limit $\hat{\Sigma}=0$ solving Eq.~\eqref{Eq:Eilenberger}  yields
 
 \begin{eqnarray}
     g_{\boldsymbol{n}_{F},1}&=&\frac{\Delta}{\sqrt{\Delta^{2}-\epsilon^{2}}}-\frac{\epsilon}{\sqrt{\Delta^{2}-\epsilon^{2}}}Ae^{\kappa x}\nonumber\\
     g_{\boldsymbol{n}_{F},2}&=&\lambda Ae^{\kappa x}\nonumber\\
     g_{\boldsymbol{n}_{F},3}&=&\frac{i\epsilon}{\sqrt{\Delta^{2}-\epsilon^{2}}}-\frac{i\Delta}{\sqrt{\Delta^{2}-\epsilon^{2}}}Ae^{\kappa x}. \label{Eq:Green function}
 \end{eqnarray}
 
  Here, $\kappa=\frac{2\sqrt{\Delta^2-\epsilon^2}}{v_F\cos(\phi)}$, where $\phi$ is the angle between $\boldsymbol{n}_{F}$ and the $x$ axis.
 $\lambda$ denotes the sign of the $x$ component of $\boldsymbol{n}_{F}$. $A$ is a constant determined by the boundary conditions. The boundary condition for an Eilenberger equation is given by [\onlinecite{zaitsev1984quasiclassical,millis1988}]
 
 \begin{equation}
   \label{eq:eilenberger-bc}
     \hat{g}_{\bar{\boldsymbol{n}}_{F}}=\hat{R}\hat{g}_{\boldsymbol{n}_{F}}\hat{R}^{\dagger},
 \end{equation}
 where $\boldsymbol{n}_{F}$ and $\bar{\boldsymbol{n}}_{F}$ are two momentum directions with the same $y$ components but opposite $x$ component. In the presence of translational invariance in the $y$ direction, electron with momentum in $\boldsymbol{n}_{F}$ direction is reflected back into an electron with momentum in $\bar{\boldsymbol{n}}_{F}$ direction. $R$ is the reflection part of the scattering matrix at the boundary, and has the form
 
 \begin{equation}
   \label{eq:reflection-matrix}
     \hat{R}=\left[\begin{array}{cc}
e^{i\theta} & 0\\
0 & e^{-i\theta}
\end{array}\right]e^{i\gamma}.
\end{equation}
 The overall phase factor $e^{i\gamma}$ does not affect the solution of the Eilenberger equation and we drop it in the rest of the paper. For a conventional $s$ wave superconductor $\theta=0$ and $A=0$, so that the quasiclassical Green function is homogeneous and there are no edge states. Solving the scattering problem for Eq.~\eqref{eq:gorkov}, we find (see Appendix~\ref{app:specular})
 \begin{equation}
   \label{eq:theta-phase}
     \theta=\arg\left(\sin\phi-iX\cos\phi\right),
 \end{equation}
where $S_{F}=\sqrt{\alpha^{2}p_{F}^{2}+h_{z}^{2}}$ and $X$ is the time reversal symmetry breaking factor defined by $X=\frac{h_z}{S_F}$. Matching the boundary conditions at $x=0$, we get

\begin{equation}
    A=\frac{\Delta\tan\theta}{\sqrt{\Delta^{2}-\epsilon^{2}}+\epsilon\tan\theta}.
\end{equation}

The density of states $N(\epsilon,x)$ is the real part of $g_3$ times the normal state density of states $1/\pi v_F$

\begin{equation}
    N(\epsilon,x)=2\frac{1}{\pi v_{F}}\Re(g_{3})=\frac{2\sqrt{\Delta^{2}-\epsilon^{2}}}{v_{F}\cos\phi}\delta(\epsilon+ \Delta\cos\theta)e^{\kappa x}.
\end{equation}

From this expression, one can see that there is a low energy qusiparticle excitation localized at the edge. This is the Majorana mode. The energy dispersion of Majorana edge states can be read out from the delta function

\begin{eqnarray}
  \epsilon=-\Delta\cos\theta&=&\frac{\text{sgn}(X)\Delta\sin\phi}{\sqrt{\sin^2\phi+\frac{h_z^2\cos^2\phi}{S_F^2}}}  \nonumber\\
  &=&\frac{\text{sgn}(X)\Delta p_y}{p_F\sqrt{1-\frac{\alpha^2(p_F^2-p_y^2)}{S_F^2}}}.
\end{eqnarray}

At low energy, the group velocity of the Majorana mode is given by

\begin{equation}
    v_M=\frac{\partial\epsilon}{\partial p_y}\approx \frac{\Delta}{p_FX},
\end{equation}
In the same method, we obtain the group velocity of the edge mode on the other edge $v_M'\approx-\frac{\Delta}{p_FX}$, indicating that the edge mode is chiral and propagates in one direction. The localization length of the zero energy Majorana mode is $l_M=v_F/\Delta$. Integrating $N(\epsilon,x)$ over $x$, we get the total density of states

 \begin{figure}[h!]
\centering
\subfigure{\includegraphics[width = 0.45\columnwidth]{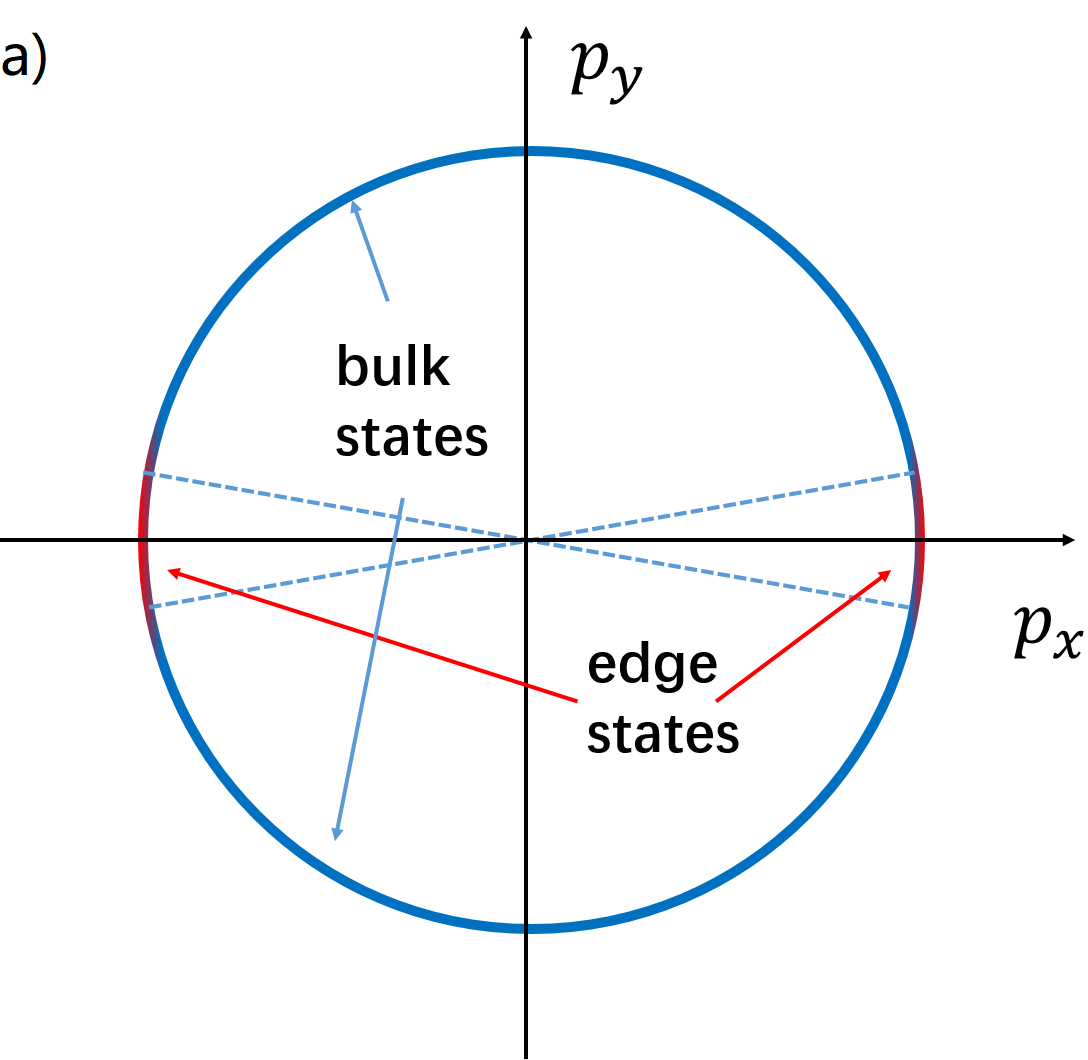}}
\subfigure{\includegraphics[width=0.45\columnwidth]{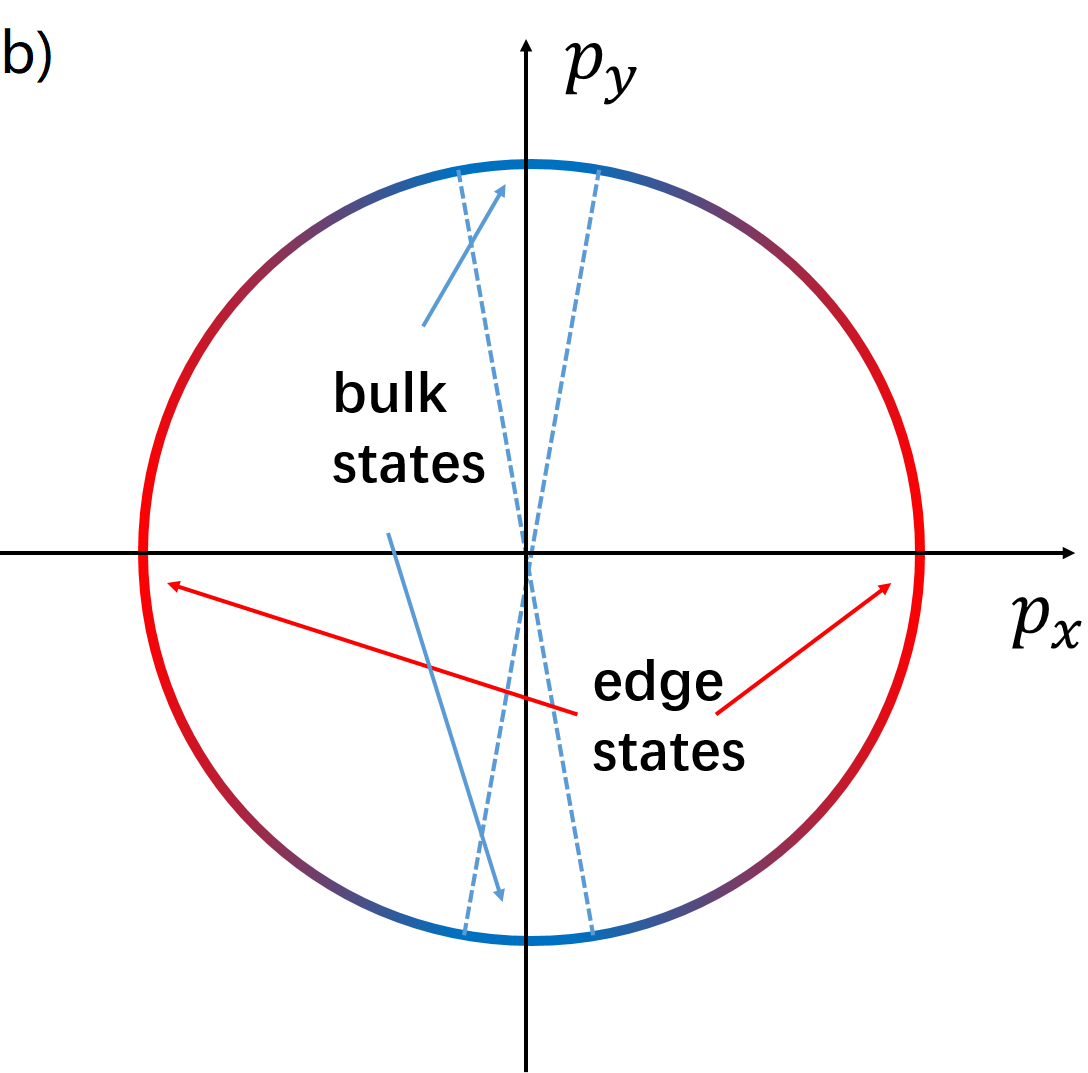}}
\caption{Sketch of Fermi surfaces. Red and blue color label edge states and bulk states, respectively. a) $X\ll 1$, there is only a small number of low energy edge states. b) $X\approx 1$, a  large portion of states on the Fermi surface contribute to edge states. \label{Fig:Fermi surface}}
\end{figure}
\begin{equation}
    N(\epsilon)=\int dx N(\epsilon,x)=\delta(\epsilon+\cos\theta),
\end{equation}
which shows that the edge mode is indeed a single channel mode. One interesting property of this chiral Majorana mode is that the number of low energy states depends on $X$. According to $\epsilon=-\Delta\cos\theta$, a low energy edge state corresponds to a large $\theta$. When $X\ll 1$, $\theta$ is finite only when $\phi$ is small. Thus there is only a small number of low energy states and the group velocity of the chiral Majorana mode is large (Fig.~\ref{Fig:Fermi surface}a).  In the opposite limit, when $X\approx 1$, $\theta$ is finite for a wide range of $\phi$, which indicates that there is a large number of low energy modes with a small group velocity (Fig. \ref{Fig:Fermi surface}b). For $X\to 1$, this model becomes similar to the spinless chiral $p$ wave superfluid [\onlinecite{tsutsumi2012edge,suzuki2016spontaneous,bakurskiy2014anomalous,sauls2011surface}]. Below we show that this property is useful for understanding the effect of strong disorder on the Majorana localization length.

\section{Effect of disorder on Majorana localization length}

In the presence of the disorder potential, we need to add the self-energy term $\hat{\Sigma}$ to the Eilenberger equation. Here we consider the weak disorder case and treat disorder potential as a perturbation. Then we can approximate the disorder self-energy as
$\hat{\Sigma}=\langle\hat{g}^0\rangle_{\boldsymbol{n}_F}/2\tau$, where $\hat{g}^0$ is the Green function without disorder given by Eq.~\eqref{Eq:Green function}. For convenience we separate the "bulk" part and the "edge" part of the Green function without disorder

\begin{equation}
    \hat{g}=\hat{g}_B+A\hat{g}_Ee^{\kappa x},
\end{equation}
where $\hat{g}_B$ is homogeous describing the bulk properties and $\hat{g}_E$ is proportional to the exponential factor $e^{\kappa x}$ describing the properties of edge states. They are given by

\begin{equation}
  \hat{g}_B=\frac{\Delta}{\sqrt{\Delta^2-\epsilon^2}}\tau_1+\frac{i\epsilon}{\sqrt{\Delta^2-\epsilon^2}}\tau_3.
\end{equation}

\begin{equation}
    \hat{g}_E=-\frac{\epsilon}{\sqrt{\Delta^2-\epsilon^2}}\tau_1+\tau_2-\frac{i\Delta}{\sqrt{\Delta^2-\epsilon^2}}\tau_3.
\end{equation}

Similarly, the self-energy can be written as

\begin{equation}
    \hat{\Sigma}=\hat{\Sigma}_B+\hat{\Sigma}_Ee^{\kappa x},
\end{equation}
where $\hat{\Sigma}_B$ is homogeneous and $\hat{\Sigma}_E$  decays exponentially away from the boundary. Since we are studying the localization length of the zero energy state, we focus on the Green function with $\boldsymbol{n}_F$ pointing to the positive $x$ direction denoted as $\hat{g}^+$. The self-energy enters the Eilenberger equation in the commutator, which is

\begin{eqnarray}
    \left[\hat{\Sigma},\hat{g}^+ \right]&=&\left[\hat{\Sigma}_B+\hat{\Sigma}_Ee^{\kappa x},\hat{g}^+_B+A\hat{g}^+_Ee^{\kappa x} \right] \nonumber\\
    &=&\left[\hat{\Sigma}_B,\hat{g}^+_B\right]+\left(\left[\hat{\Sigma}_E,\hat{g}^+_B\right]+A\left[\hat{\Sigma}_B,\hat{g}^+_E\right]\right)e^{\kappa  x}\nonumber \\&+&A\left[\hat{\Sigma}_E,\hat{g}^+_E\right]e^{2\kappa x}.\label{Eq:self-energy}
\end{eqnarray}
Since we are only interested in the Majorana localization legnth, we focus on the Green function far away from the boundary, where $\hat{g}_E$ and $\Sigma_E$ can be treated as perturbations. Thus we can drop the third term on the right hand side of Eq.~\eqref{Eq:self-energy} which is a higher order perturbation. Note that in the second term on the right hand side of Eq.~\eqref{Eq:self-energy}, $A$ is divergent, such that we can ignore the $\hat{\Sigma}_E\hat{g}_B$ term. Then it can be seen that $\hat{\Sigma}_E$ does not appear in the Eilenberger equation. The disorder self-energy has only a bulk contribution, which in the weak disorder limit is given by

\begin{equation}
    \hat{\Sigma}\approx\hat{\Sigma}_B=\left(\frac{1}{2}\hat{g}_B+\frac{X^2}{2}\tau_3\hat{g}_B\tau_3\right)/\tau.\label{Eq:Bulk self-energy}
\end{equation}

Substituting Eq.~\eqref{Eq:Bulk self-energy} into Eq.~\eqref{Eq:Eilenberger}, we obtain the Eilenberger equation in the presence of weak disorder

\begin{equation}
       \boldsymbol{v}_{F}'\cdot\hat{\boldsymbol{\nabla}}\hat{g}_{\boldsymbol{n}_{F}}=\left[\hat{g}_{\boldsymbol{n}_{F}},i\epsilon\tau_{3}+\Delta'\tau_{1}\right].\label{Eq:Eilenberger2}
 \end{equation}

This Eilenberger equation has exactly the same structure as that in the clean case but with a renormalized Fermi velocity and pairing potential, which are given by

\begin{eqnarray}
    v_F'&=&\frac{v_F}{1+\frac{1}{2\tau \Delta}\left(1+X^2\right)}\nonumber\\
    \Delta'&=&\frac{\Delta \left[1+\frac{1}{2\tau \Delta}\left(1-X^2\right)\right]}{1+\frac{1}{2\tau \Delta}\left(1+X^2\right)},
\end{eqnarray}
 It can be seen that both Fermi velocity and pairing potential are reduced by disorder. The Majorana localization length is thus

\begin{equation}
    l_M=\frac{v_F'}{\Delta'}=\frac{v_F}{\Delta\left[1+\frac{1}{2\tau \Delta}\left(1-X^2\right)\right]}.\label{Eq:localization length}
\end{equation}
 Since $X^2<1$, $l_M$ is always smaller than $l_{M0}=v_F/\Delta$, which is the Majorana localization length in the clean case. Weak disorder thus reduces the Majorana localization length for any SO coupling strength and exchange field. This effect is opposite to that in one dimension, where weak disorder usally increases the Majorana localization length [\onlinecite{motrunich2001griffiths,brouwer2011probability,lobos2012interplay,pientka2013signatures,huse2013localization,adagideli2014effects,hui2014generalized}]. The main difference between 2D and 1D systems is that in two dimensions there are many states near the Fermi energy and only a few of them contribute to the Majorana edge states. Hence, at weak disorder the disorder self-energy has only a bulk contribution. However, in one dimension, there are only two channels near the Fermi energy, both of which contribute to the Majorana end states. Thus, the edge contribution to the self-energy has a large impact on the Majorana localization length.

\section{Majorana localization length for arbitrary disorder strength}

\begin{figure}[h!]
\centering
\includegraphics[width = 0.9\columnwidth]{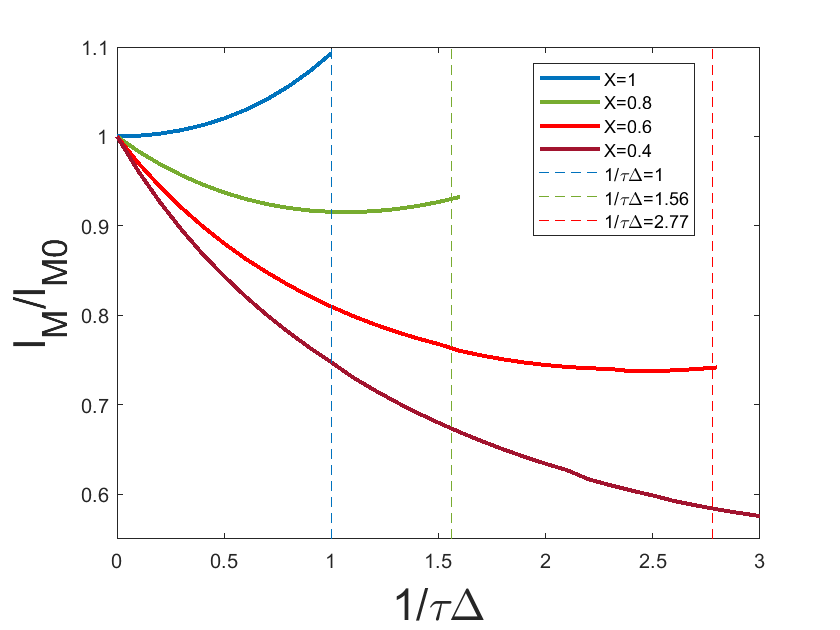}

\caption{Majorana localization length $l_M$ versus disorder strength for different time reversal symmetry breaking factors $X=1,0.8,0.6,0.4$. The vertical dashed lines label the gap closing points. At weak disorder, $l_M$ decreases with increasing $1/\tau$ while at large disorder $l_M$ increases with increasing $1/\tau$. The critical disorder $1/\tau_c$ is much smaller than pairing $1/\tau_c\ll \Delta$ for $X=1$ and increases when increasing $X$. Here Majorana localization length is normalized by $l_{M0}$ and the disorder strength is normalized by the order parameter $\Delta$. }\label{Fig:localization length}
\end{figure}

In order to obtain the Majorana localization length for an arbitrary disorder strength, we numerically solve Eq.~(\ref{Eq:Eilenberger}) (Appendix~\ref{app:numerical}). Here we use an exponential function $\text{DOS}=Ae^{-\kappa x}$ to fit the tail of the spatial dependent density of states and the Majorana localization length $l_M$ is defined as $l_M=2/\kappa$. The result is shown in Fig. \ref{Fig:localization length}. It can be seen that weak disorder decreases $l_M$ for $X=0.8,0.6,0.4$. Increasing the disorder strength, the Majorana localization length starts to increase after the disorder strength reaches the critical value $1/\tau_c$. The critical disorder strength depends on $X$. In particular $1/\tau_c$ goes to zero when $X$ approaches 1 and increases monotonically with decreasing $X$. To understand the behavior of $\tau_c$, we note that the increase of $1/\tau_c$ is caused by the edge self-energy $\hat{\Sigma}_E$. For small $X$, as  mentioned above, the number of edge states is small (Fig. \ref{Fig:Fermi surface}), and thus a large disorder strength is required to increase $l_M$. Thus the critical disorder is large.        
Note that near the gap closing point $1/\tau=\Delta/X^2$(Appendix.~\ref{app:gap closing point}), the Majorana localization length is finite unlike in  the 1D case where the Majorana localization length is divergent near the gap closing point. We explain this  in Appendix.~\ref{app:localization length}.

\section{Discussion and conclusion}

In conclusion, we use quasiclassical theory to  study the effect of disorder on the Majorana localization in a two dimensional topological superconductor. We find the nonmonotonic behavior of the Majorana localization length $l_M$ as a function of disorder strength. We show that weak disorder decreases $l_M$ while strong disorder increases  it. The critical disorder strength $1/\tau_c$ where $\frac{dl_M}{d\tau}\big|_{\tau=\tau_c}=0$ depends on the time reversal symmetry breaking factor $X=h_z/S_F$. $1/\tau_c$ tends to zero when $|h_z|\ll \alpha p_F$ and increases when reducing $X$.

The fact that disorder can decrease Majorana localization length was first reported in one dimensional multi-channel superconductors [\onlinecite{haim2019benefits}]. In our work, the physics is different from [\onlinecite{haim2019benefits}]. In our case, the chemical potential only cuts one band in the normal state and the decreased Majorana localization length is attributed to the renormalized Fermi velocity. Although in this work we study a specific model, our results are valid in any two dimensional gapped topological superconductors. This is because the renormalization of the Fermi velocity is universal in two dimensional superconductors, but the time reversal symmetry breaking factor $X$ has different expressions in different models [\onlinecite{he2017chiral}] depending on the type of SO coupling and the direction of the magnetic field.

\begin{acknowledgments}
This work is supported by the Academy of Finland project HYNEQ (Project no. 317118).
\end{acknowledgments}

\appendix
\section{Basis of the projected Eilenberger equation}
\label{app:basis}
Since the chemical potential cuts only the lower band, we can ignore the high energy band and project the Eilenberger equation onto the lower band. The eigenvector of the lower band used here is 

\begin{equation}
|\psi_{-,e}\rangle=\begin{pmatrix}\begin{array}{cccc}
\alpha p_{F}e^{i\phi'/2}, & (S_{F}+h_{z})e^{-i\phi'/2}, & 0, & 0\end{array}\end{pmatrix}^{\text{T}}/N
\end{equation}
\begin{equation}
|\psi_{-,h}\rangle=\begin{pmatrix}\begin{array}{cccc}
0, & 0, & (S_{F}+h_{z})e^{i\phi'/2}, & \alpha p_{F}e^{-i\phi'/2}\end{array}\end{pmatrix}^{\text{T}}/N,
\end{equation}
where $|\psi_{-,e}\rangle$ and $|\psi_{-,h}\rangle$ are electron and hole parts of the eigenvector, respectively. $N$ is the normalization factor $N=\sqrt{2S_F(S_F+h_z)}$. $\phi'$ is the angle between the momentum direction and the $y$ axis.

\section{Specular hard-wall scattering}

\label{app:specular}

The quasiclassical boundary condition~\eqref{eq:eilenberger-bc}
is expressed in terms of the scattering matrix of the interface [\onlinecite{millis1988}].
To find it, we solve here the specular hard-wall
scattering problem for Eq.~\eqref{eq:gorkov} in the normal state,
for the $2\times2$ electron and hole blocks $H_{\tau}=(\epsilon\tau_z-\mu-\hat{G}_0^{-1})\rvert_{\tau_z\mapsto\tau=\pm1}$.
The bulk material resides at $x<0$ and is terminated by the boundary at $x=0$.
Due to the exchange field in the bulk, the reflection phase is not necessarily
the same for electrons and holes, and needs to be calculated explicitly.

We assume $\mu$ is such that there is a single Fermi surface, on
the lower helical band. The scattering wave function at $x\le0$ is
\begin{align}
  \Psi_{\tau}(x) = e^{ik_{\tau}x}\chi_{\mathrm{i},\tau} + r_{\tau} e^{-ik_{\tau}x}\chi_{\mathrm{o},\tau} + c_{\tau} e^{\kappa_{\tau} x}\chi_{\mathrm{ev},\tau}
  \,,
\end{align}
where $\chi_{\mathrm{i}, \mathrm{o}, \mathrm{ev}}$ are
2-element spinors satisfying $H_{\tau}(k_x,k_y)\chi_\tau=\tau\epsilon\chi_\tau$ at
$k_x=+k_{\tau}$, $k_x=-k_{\tau}$, and the evanescent wave vector
$k_x=-i\kappa_{\tau}$, respectively.  Here,
$\chi_{\mathrm{i},\mathrm{o}}$ can be normalized to
$\Vert\chi\Vert^2=\mathrm{const.}$ as they carry the same
current. Then, $r_{\tau}$ is the reflection amplitude.

We can note that $H_{\tau}(-k,k_y)=H_{\tau}(k,k_y)^*$ and that
$H_{\tau}(-i\kappa,k_y)$ is real-valued, so that we can choose
$\chi_{\mathrm{o},\tau}=\chi_{\mathrm{i},\tau}^*$ and $\chi_{\mathrm{ev},\tau}$ real.

The hard-wall boundary condition $\Psi_{\tau}(0)=0$ results to the
reflection amplitude
\begin{align}
  r_{\tau}
  =
  -\frac{\det(\chi_{\mathrm{i},\tau},\chi_{\mathrm{ev},\tau})}{\det(\chi_{\mathrm{o},\tau},\chi_{\mathrm{ev},\tau})}
  =
  -\frac{\det(\chi_{\mathrm{i},\tau},\chi_{\mathrm{ev},\tau})}{\det(\chi_{\mathrm{i},\tau},\chi_{\mathrm{ev},\tau})^*}
  \,.
\end{align}
Hence, $r_{\tau}=-\exp[2i\arg\det(\chi_{\mathrm{i},\tau},\chi_{\mathrm{ev},\tau})]$.

The quasiclassical reflection amplitude is evaluated at the Fermi
surface, $\epsilon=0$. There, $k_-=k_+$ and $\kappa_-=\kappa_+$, and
using $H_{-}(k_x)=\sigma_1 H_+(k_x^*)^* \sigma_1$ we can choose
$\chi_{\mathrm{i},-}=\sigma_1 \chi_{\mathrm{i},+}^*$ and
$\chi_{\mathrm{ev},-}=\sigma_1\chi_{\mathrm{ev},+}\rvert_{-\kappa}$ where
$\chi_{\mathrm{ev},+}\rvert_{-\kappa}$ is the real evanescent spinor at
$k_x=+i\kappa$. Then, $r_+r_-^*=e^{2i\theta}$ where
\begin{align}
  \label{eq:theta-eq}
   \theta&=\arg z\,,
   \\
   z&=\det(\chi_{\mathrm{i},+},\chi_{\mathrm{ev},+})\det(\chi_{\mathrm{i},+},\chi_{\mathrm{ev},+}\rvert_{-\kappa})
   \,.
\end{align}
In the wave function basis used here (see App.~\ref{app:basis}),
\begin{align}
  \label{eq:chi-spinors}
  \chi_{\mathrm{i},+}\propto\begin{pmatrix} \alpha k_F e^{i\phi'/2} \\ (S_F + h_z)e^{-i\phi'/2} \end{pmatrix}
  \,,
  \;
  \chi_{\mathrm{ev},+}\propto\begin{pmatrix} \alpha (k_y + \kappa) \\ S_F' + h_z \end{pmatrix}
  \,,
\end{align}
where $\phi'=\frac{\pi}{2}-\phi$, $k_{Fx}=k_F\sin\phi'$,
$k_y=k_F\cos\phi'$, $k_F^2=2m(S_F+\mu)$, $\kappa^2=k_y^2 - 2m(S_F' + \mu)$, and $S_F'=2m\alpha^2-S_F=-S_F+(S_F^2-h_z^2)/(S_F+\mu)$.  A mechanical if long calculation
making these substitutions gives:
\begin{align}
  z = 4m \alpha^2(\mu-h_z)S_F(S_F-S_F')\bigl[\sin\phi - i\frac{h_z}{S_F}\cos\phi\bigr]
  \,.
\end{align}
From this and Eq.~\eqref{eq:theta-eq}, we find Eq.~\eqref{eq:theta-phase}.

\section{Zaitsev's boundary conditions}

Once the reflection matrix is known, we use the decoupling of the
equations for the slowly varying quasiclassical parts from the
fast-oscillating parts of the Green function derived in
Refs.~\onlinecite{zaitsev1984quasiclassical,millis1988}.  Because the
problem here involves a projection to the lower band which complicates
the discussion, we outline here for completeness how it can be
handled.  We also limit the discussion to the fully reflective interface,
where the problem becomes simpler.

We consider the same setup as in App.~\ref{app:specular}, with
interface at $x=0$, but with Hamiltonian at $x<0$
slowly varying on a length scale $\lambda\gg{}\kappa^{-1}$, $k_{Fx}^{-1}$.
When $|x-x'|,-x,-x'\gg\kappa^{-1}$, the Green function Ansatz, for a
fixed $k_y$, is
\begin{gather}
  \label{eq:Gansatz}
  \hat{G}_1(x,x')
  =
  \sum_{ab=\pm} e^{i k_{Fx} (ax - bx')} \hat{C}_{ab}(x,x')
  \,,
  \\
  \hat{C}_{ab}(x,x')
  =
  \sum_{\tau,\tau'=\pm}
  |\psi_{a,-,\tau}\rangle \langle\psi_{b,-,\tau'}| (C_{ab})_{\tau\tau'}(x,x')
  \,,
\end{gather}
where $C_{ab}(x,x')=\theta(x-x')C_{ab}^>(x,x')+\theta(x'-x)C^<_{ab}(x,x')$ and $C_{ab}^{>/<}$ are slowly
varying amplitudes. Moreover,$\lvert\psi_{a,-,\tau}\rangle$
 are the lower-band
null vectors, satisfying $\hat{H}_0(ak_{Fx},k_y)\lvert\psi_{a,-,\tau}\rangle=0$ for the normal-state bulk Hamiltonian $H_0$ which is
block-diagonal in the Nambu index $\tau$.

Andreev approximation in the Gor'kov equation for
$\hat{G}^{-1}=\epsilon\tau_3 - \hat{H}(x,-i\partial_x)$ with slowly
varying $\hat{H}(x)$, and projection to the lower band gives, for
$x\ne{}x'$,
\begin{align}
  0
  &\simeq
  \langle\psi_{a,-,\tau}|[\epsilon\tau_3 - \hat{H}(x,a k_{Fx})]\hat{C}_{ab}
  |\psi_{b,-,\tau'}\rangle
  \\\notag&
  \qquad
  + i
  \langle \psi_{a,-,\tau}|v_x (\partial_x\hat{C}_{ab})
  |\psi_{b,-,\tau'}\rangle
  \\
  \label{eq:andreev-left}
  &=
  \bigl([\epsilon\tau_3 - \tilde{H}(x,a k_{Fx})]C_{ab}
  + i a v_x \partial_x C_{ab}
  \bigr)_{\tau\tau'}
  \,,
\end{align}
and similarly for the adjoint equation,
\begin{align}
  \label{eq:andreev-right}
  0
  &\simeq
  \bigl(C_{ab}[\epsilon\tau_3 - \tilde{H}(x,b k_{Fx})]
  - i b \partial_{x'} C_{ab} v_x
  \bigr)_{\tau\tau'}
  \,.
\end{align}
Here,
$(v_x)_{\tau\tau'} = \langle\psi_{-,a,\tau}|(k_{Fx}/m +
a\alpha\sigma_2)|\psi_{-,a,\tau'}\rangle = \delta_{\tau\tau'}[1 -
\frac{m\alpha^2}{S_F}]\frac{k_{Fx}}{m}=\delta_{\tau\tau'}v_F\sin\phi'$ is diagonal, and $\tilde{H}$ is
the projected Hamiltonian.  Hence, for $x$ away from the interface and
$\lambda\gg\delta\gg\kappa^{-1}$, $C_{++}(x,x\pm\delta)$,
$C_{--}(x,x\pm\delta)$ follow the quasiclassical Eilenberger equation.

When $|x-x'|\gg\kappa^{-1}$ but either $x$ or $x'$ is close to the interface at $x=0$,
the
evanescent state
$\hat{H}_0(-i\kappa,k_y)\lvert\psi_{\mathrm{ev},\tau}\rangle=0$ also
has a finite amplitude:
\begin{gather}
  \hat{G}_2
  =
  \hat{G}_1
  +
  \begin{cases}
  \sum_{b=\pm}
  e^{\kappa x - i b k_{Fx} x'}
  \hat{C}_{0b}
  \,,
  &
  \text{for $x>x'$,}
  \\
  \sum_{a=\pm}
  e^{iak_{Fx} x+\kappa x'}
  \hat{C}_{a0}
  \,,
  &
  \text{for $x<x'$,}
  \end{cases}
  \\
  \hat{C}_{a0}=\sum_{\tau\tau'}
  (C_{a0})_{\tau\tau'}
  |\psi_{a,-,\tau}\rangle \langle\psi_{\mathrm{ev},\tau'}|
  \,,
  \\
  \hat{C}_{0b}=\sum_{\tau\tau'}
  (C_{0b})_{\tau\tau'}
  |\psi_{\mathrm{ev},\tau}\rangle \langle\psi_{b,-,\tau'}|
  \,.
\end{gather}
%
The Ansatz by construction satisfies $(H_0G_2)(x,x')=(G_2H_0)(x,x')=0$
when $C_{ab}^{>/<}$ are constant.
It satisfies also the boundary conditions $\hat{G}_2(0,x')=\hat{G}_2(x,0)=0$ if
\begin{align}
  0  
  &=
  (C_{+,b}^>)_{\tau\tau'}\lvert \psi_{+,-,\tau}\rangle
  +
  (C_{-,b}^>)_{\tau\tau'}\lvert \psi_{-,-,\tau}\rangle
  \\\notag&\qquad
  +
  (C_{0,b}^>)_{\tau\tau'}\lvert \psi_{\mathrm{ev},\tau}\rangle
  \,,
  \\
  0
  &=
  (C_{a,+}^<)_{\tau\tau'}
  \langle \psi_{+,-,\tau'}\rvert
  +
  (C_{a,-}^<)_{\tau\tau'}
  \langle \psi_{-,-,\tau'}\rvert
  \\\notag&\qquad
  +
  (C_{a,0}^<)_{\tau\tau'}
  \langle \psi_{\mathrm{ev},\tau'}\rvert
  \,.
\end{align}
This is the scattering problem solved in App.~\ref{app:specular}
above. The solution gives the boundary conditions $C_{++}^>=\hat{R}^\dagger{}C_{-+}^>$,
$C_{--}^>=\hat{R}C_{+-}^>$, $C_{++}^<=C_{+-}^<\hat{R}$,
$C_{--}^<=C_{-+}^<\hat{R}^\dagger$ where
$\hat{R}=\mathop{\mathrm{diag}}(r_+, r_-)$ is the reflection
matrix~\eqref{eq:reflection-matrix}.  Note that the results here are
more limited than in [\onlinecite{millis1988}], as we assume the special case of a nontransparent and sharp interface, where the normal-state Hamiltonian stays constant up to the interface.

Writing the Green function around $x=x'$ also needs inclusion of
additional terms $\propto{}e^{\mp\kappa(x-x')}$.  The exact Green
function is continuous at $x=x'$ with the jump condition
$[\partial_x\hat{G}]_{x=x'-0^+}^{x=x'+0^+}=2m$. For the
Ansatz at $x=x'$, this implies continuity of the drone amplitudes,
$C_{+-}^<(x,x)=C_{+-}^>(x,x)$, $C_{-+}^<(x,x)=C_{-+}^>(x,x)$, as they
are the only components oscillating as $e^{\pm2ik_{Fx}x}$.
The relations between $C_{aa}^<(x,x)$ and $C_{aa}^>(x,x)$ are more complicated,
but are not necessary to find the reflective boundary condition.
Together with the scattering boundary conditions, this implies that close to the interface (for $\lambda\gg{}|x|,|x'|\gg\kappa^{-1}$),
$C^>_{++}=\hat{R}^\dagger{}C_{--}^<\hat{R}$,
$C^<_{++}=\hat{R}^\dagger{}C_{--}^>\hat{R}$.

The remaining problem is to relate $C_{ab}^{>/<}$ to the quasiclassical Green
function. To do this, we move Eq.~\eqref{eq:Gansatz} to the Wigner
representation assuming slowly varying $C_{ab}$, and drop the $\pm2k_{Fx}$
drone amplitudes:
\begin{align}
  \label{eq:Gwigner}
  \hat{G}(k_x;x)
  &\simeq
  \sum_{a=\pm}
  \Bigl(
  \frac{\hat{C}_{aa}^>(x,x)}{\eta - i(ak_{Fx} - k_x)}
  +
  \frac{\hat{C}_{aa}^<(x,x)}{\eta + i(ak_{Fx} - k_x)}
  \Bigr)
  \,,
\end{align}
where $\eta\to0^+$.
The quasiclassical Green function $\hat{g}$ is obtained by integrating
over $\xi=v_F(k - k_F)=v_F\delta k$ in the vicinity of the Fermi surface after fixing the momentum direction so that $k_x =k\sin\phi'$ and $k_y=k\cos\phi'$.  Because
$k_{Fx}=\sqrt{k_F^2 - k_y^2}=\sqrt{k_F^2-k^2\cos^2\phi'}$ also
depends on $k$, linearizing around $k\approx{}k_F$ in~\eqref{eq:Gwigner} gives
\begin{align}
  ak_{Fx} - k_x
  \simeq
  (a-a')(k_F+\delta k)|\sin\phi'| - \frac{a\delta k}{|\sin\phi'|}
  \,,
\end{align}
where $a'=\mathop{\mathrm{sgn}}\sin\phi'$.
Hence, we have for $\hat{g}(x,\phi')$:
\begin{align}
  \hat{g}(x,\phi')
  &=
  \frac{i}{\pi}
  \int v_F d(k-k_F)\, P_- \hat{G}(k_x;x) P_-^\dagger
  \notag
  \\
  &\simeq
  iv_F |\sin\phi'|[
  C^>_{a'a'}(x,x)
  +
  C^<_{a'a'}(x,x)
  ]
  \,,
\end{align}
where $P_-$ is the projector to the lower band, and only the $\delta$-function
parts are included in the $\xi$-integration.  The boundary
conditions for $C_{aa}^{>/<}$ now imply a Zaitsev boundary condition
for $\hat{g}$:
\begin{align}
  \hat{g}(x=0,-\phi')
  &=
  \hat{R}
  \hat{g}(x=0,\phi') 
  \hat{R}^\dagger
  \,,
\end{align}
for $\sin\phi'>0$, and we find Eq.~\eqref{eq:eilenberger-bc}.

The quasiclassical approach neglects a fast-oscillating part,
which contributes a $\cos(2k_Fx)$ term in the DOS. However, we do not
need to consider it in the problem with a single interface, as the
equations for the slowly varying $\hat{g}$ are decoupled from the
fast part.  Similar decoupling was previously obtained also from a
different approach, explicitly for the 1D Majorana problem with a
semi-infinite disordered bulk and a single interface
[\onlinecite{hui2014generalized}].  However, interference effects
e.g. between multiple interfaces are not captured in the
quasiclassical approach [\onlinecite{zaitsev1984quasiclassical}].
This includes e.g. the $k_fL$ oscillation of the energy level of
overlapping Majorana end states [\onlinecite{stanev2014}].

\section{Numerical calculation}
\label{app:numerical}
We solve the Eilenberger equation numerically by using the simple iteration method. We first calculate the Green function $\hat{g}_1$ in the absence of disorder. Then we substitute the disorder self-energy $\hat{\Sigma}_1=\hat{g}_1/\tau$ back into the Eilenberger equation and obtain another Green function $\hat{g}_2$. We repeat this process for several times until the difference between $\hat{g}_n$ and $\hat{g}_{n+1}$ is smaller than 0.001.  
\section{Gap closing point}
\label{app:gap closing point}
In this appendix, we show how to find the gap closing point for both 1D and 2D cases. We can write the bulk Green function as $\hat{g}_B=g_{B,1}\tau_1+g_{B,2}\tau_2+g_{B,3}\tau_3$, which is independent of position and momentum direction. In 1D, $\hat{g}_B$ satisfies the Eilenberger equation

\begin{equation}
    [i\epsilon\tau_3+\Delta\tau_1+X^2\tau_3\hat{g}_B\tau_3/2\tau,\hat{g}_B]=0.\label{eq:Bulk equation 1d}
\end{equation}

In 2D, the bulk Eilenberger equation is given by

\begin{equation}
    [i\epsilon\tau_3+\Delta\tau_1+\hat{g}_B/2\tau+X^2\tau_3\hat{g}_B\tau_3/2\tau,\hat{g}_B]=0.\label{eq:Bulk equation 2d}
\end{equation}
Note that Eq.~\eqref{eq:Bulk equation 1d} is equivalent to Eq.~\eqref{eq:Bulk equation 2d} because $[\hat{g}_B/2\tau,\hat{g}_B]=0$. Setting $\epsilon=0$, we get two solutions to Eq.~\eqref{eq:Bulk equation 1d} 

\begin{equation}
    g_{B,2}=0,\quad g_{B,3}=0,\quad g_{B,1}=1
\end{equation}
or
\begin{equation}
    g_{B,2}=0,\quad g_{B,3}=\sqrt{1-\Delta^2\tau^2/X^4},\quad g_{B,1}=\Delta\tau/X^2.
\end{equation}
These two solutions coincide at $1/\tau=\Delta/X^2$. Making use of the "boundary conditions" $\hat{g}_B(1/\tau=0)=\tau_1$, $\hat{g}_B(1/\tau\rightarrow+\infty)=\tau_3$ [\onlinecite{lu2019proximity}], we find the physical solution, which is for $1/\tau<\Delta/X^2$
\begin{equation}
    g_{B,2}=0,\quad g_{B,3}=0,\quad g_{B,1}=1
\end{equation}
and for $1/\tau<\Delta/X^2$
\begin{equation}
     g_{B,2}=0,\quad g_{B,3}=\sqrt{1-\Delta^2\tau^2/X^4},\quad g_{B,1}=\Delta\tau/X^2 .
\end{equation}
Therefore the gap closing point is $1/\tau=\Delta/X^2$.

\section{Majorana localization length near the gap closing point}
\label{app:localization length}
\subsection{One dimensional case}
In one dimension, the Eilenberger equation is given by
\begin{equation}
    v_F\nabla\hat{g}_{\lambda}=\left[\hat{g}_{\lambda},i\epsilon\tau_3+\Delta\tau_1+\hat{\Sigma}_{\lambda}\right],\label{eq:Eilenberger1d}
\end{equation}
where $\lambda=+/-$ corresponds to right/left going electrons. The disorder self-energy is given by $\hat{\Sigma}_{\lambda}=X^2\tau_3\hat{g}_{-\lambda}\tau_3/2\tau$. For convenience, we write the Green function as
\begin{equation}
    \hat{g}_{\lambda}=\hat{g}_{B,\lambda}+\hat{g}_{E,\lambda}
\end{equation}
where $\hat{g}_{B,\lambda}$ and $\hat{g}_{E,\lambda}$ are bulk and edge Green function, respectively. Far away from the boundary, $\hat{g}_{E}$ is much smaller than $\hat{g}_B$ and can be treated as a perturbation. Thus we can expand Eq.~\eqref{eq:Eilenberger1d} up to the first order in $\hat{g}_{E}$. The 0th order terms are gone because they just give the bulk Eilenberger equation. The first order terms are given by

\begin{eqnarray}
    v_F\nabla\hat{g}_{E,\lambda}&=&\left[\hat{g}_{E,\lambda},\Delta\tau_1+X^2\tau_3\hat{g}_{B,-\lambda}\tau_3/2\tau\right]\nonumber \\
    &+&\left[\hat{g}_{B,\lambda},X^2\tau_3\hat{g}_{E,-\lambda}\tau_3/2\tau\right].\label{eq:Eilenberger 1st order}
\end{eqnarray}
Here we have already set $\epsilon=0$. Before the gap closes the bulk Green function is just $\hat{g}_{B,\lambda}=\tau_1$. We also notice that $\hat{g}_{E,+}$ and $\hat{g}_{E,-}$ have the relation $\hat{g}_{E,+,1}=\hat{g}_{E,-,1}$, $\hat{g}_{E,+,2}=-\hat{g}_{E,-,2}$, $\hat{g}_{E,+,3}=\hat{g}_{E,-,3}$ for the components $\hat{g}_{E,\lambda}=\sum_{j=1}^3 \hat{g}_{E,\lambda,j} \tau_j$. Thus  Eq.~\eqref{eq:Eilenberger 1st order} can be simplified as

\begin{equation}
    v_F\nabla\hat{g}_{E,\lambda}=\left[\hat{g}_{E,\lambda},(\Delta-X^2/\tau)\tau_1\right]
    \label{eq:e4}
\end{equation}
It can be seen that the effective pairing is reduced to $\Delta-X^2/\tau$. The Majorana localization length is given by $l_M=v_F/(\Delta-X^2/\tau)$, which is divergent at the gap closing point $1/\tau=\Delta/X^2$. Our result is at odds with the numerical study in Ref. [\onlinecite{hui2014generalized}], which finds $l_M=v_F(\Delta-1/\tau)^{-0.84}$. However, it is consistent with the result from the transfer matrix method in Ref.~[\onlinecite{haim2019benefits}] despite the fact that this paper disregards the edge contribution to the self-energy for strong disorder in the Born approximation approach.

\subsection{Two dimensional case}
In two dimensions, the Eilenberger equation is given by Eq.~\eqref{Eq:Eilenberger}. Using the same method as in the one dimensional case, we arrive at

\begin{equation}
    v_F\nabla\hat{g}_{E,+}=\left[\hat{g}_{E,+},(\Delta+1/2\tau-X^2/2\tau)\tau_1\right]+\left[\tau_1,\hat{\Sigma}_{E}\right],
    \label{eq:e5}
\end{equation}
where $\hat{g}_{E,+}$ is the edge Green function with relative momentum pointing to the positive $x$ direction.  If we would assume $\hat{\Sigma}_E=(1+X^2)\hat{g}_{E,+}/2\tau$, Eq.~\eqref{eq:e5} would be simplified as
\begin{equation}
    v_F\nabla\hat{g}_{E,+}=\left[\hat{g}_{E,+},(\Delta-X^2/\tau)\tau_1\right].
    \label{eq:e6}
\end{equation}
Equation \eqref{eq:e6} is almost the same as Eq.~\eqref{eq:e4}. At the gap closing point $1/\tau=\Delta/X^2$, the effective pairing is reduced to 0 and the Majorana localization length  diverges. However, in practice $\hat{\Sigma}_E$ is smaller than $(1+X^2)\hat{g}_{E,+}/2\tau$ and is not large enough to reduce the effective pairing to zero. Therefore the Majorana localization length is finite at the gap closing point.
\bibliography{ref}

\end{document}